\documentstyle[prl,twocolumn,aps]{revtex}
\RequirePackage{epsfig}
\def\Journal#1#2#3#4{{#1} {\bf #2}, #3 (#4)}
\def\PR{\em Phys. Rev.}
\def\PRL{\em Phys. Rev. Lett.}
\def\PRA{{\em Phys. Rev.} A}
\def\JMP{\em J. Math. Phys.}

\def\Science{\em Science}
\def\JPB{\em J. Phys. B}
\def\EPJD{{\em Eur. Phys. J.} D}
\def\RMP{\em Rev. Mod. Phys.}
\begin{document}
\draft
\title{Breakdown of time-dependent mean-field theory for a\\
       one-dimensional condensate of impenetrable bosons}
\author{M. D. Girardeau$^{1,2}$ and E. M. Wright$^2$}
\address{$^1$Institute of Theoretical Science, University of Oregon, 
Eugene, OR 97403\\
	$^2$Optical Sciences Center and Department of Physics,
University of Arizona, Tucson, AZ 85721}
 \date{\today}
\maketitle
\begin{abstract}
We show that the time-dependent nonlinear Schr\"odinger equation of
mean-field theory has limited utility for a one-dimensional condensate of
impenetrable bosons.  Mean-field theory with its associated order parameter
predicts interference between split condensates that are recombined,
whereas an exact many-body treatment shows minimal interference.
\end{abstract}
Mean-field theory (MFT) has proven remarkably successful at predicting both
the static and dynamic behavior of Bose-Einstein condensates (BECs)
in weakly-interacting atomic vapors \cite{DalGioPit99},
including the ground state properties \cite{BayPet96,KagShlWal96}, the
spectrum of collective excitations \cite{EdwDodCla96,Str96},
four-wave mixing \cite{GolPlaMey95,DenHagWen99},
matter-wave solitons \cite{MorBalBur97,ReiCla97},
and interference between BECs
\cite{NarWalSch96,JavWil97,LegSol98,WriWonCol97}.
The basic notion underlying MFT
is of a macroscopic wave function \cite{PenOns56,Yan62}, or
order parameter, which defines the spatial mode into which a
significant fraction of the atoms condense below the critical temperature.
The macroscopic wave function typically obeys a nonlinear Schr\"odinger
equation (NLSE), the Gross-Pitaevskii equation, and is most suitable
for dilute Bose gases.  But the success of MFT is
not assured in all cases. For example, in one-dimension (1D) a spatially 
homogeneous ideal gas in its many-body ground state exhibits complete 
BEC into the lowest single-particle state, 
but no BEC at any nonzero temperature. Furthermore, previous exact analysis 
\cite{Gir60,Gir65} of a gas of impenetrable or hard core bosons in 1D by one 
of us (MG) and its extension by Lenard \cite{Len66} and Vaidya and Tracey 
\cite{VTr79} have shown that in the many-body ground state the
occupation of the lowest single-particle state is of order $\sqrt{N}$ where
$N$ is the total number of atoms, in contrast to $N$ for usual BEC.
Nevertheless, since $N\gg 1$ and the momentum distribution has a sharp peak
in the neighborhood of zero momentum \cite{VTr79,Ols98}, this system shows
some coherence effects.  Moreover, Olshanii \cite{Ols98}
has shown theoretically that such a 1D gas of impenetrable bosons can be
realized at sufficiently low temperatures and densities in thin atom
waveguides, and there is also considerable experimental effort towards
realizing atom waveguides for producing 1D atom clouds
\cite{HinBosHug98,ThyWesPre99}, a new and
fertile system for studying the physics of condensed systems.

Our goal in this Letter is to show that mean-field theory breaks down
in the analysis of the dynamics of 1D atom clouds, in that it
predicts interference effects that are absent in the exact theory.  In
particular, we have recently analyzed many-body solitons in a 1D
gas of bosons using the time-dependent generalization of the
Fermi-Bose mapping \cite{GirWri00a}, which maps a 1D gas of hard core
bosons to a gas of free fermions \cite{Gir60,RojCohBer99},
and Kolomeisky {\it et al.} \cite{KolNewStr00}
have proposed a corresponding NLSE with
a quartic nonlinearity to extend the usual mean-field
theory for 1D atom clouds.  For a harmonic trap the ground-state
density profiles from their theory show excellent agreement with the
exact many-body results (see Fig. 1 of their paper).  The key question, then, 
is whether this extended NLSE can be used in all circumstances. 
To address this issue we examine the problem of a 1D atomic cloud in
the ground state of a harmonic trap that is split by a blue-detuned
laser and recombined, both using an exact many-body
treatment based on the Fermi-Bose mapping and the approximate NLSE:
the NLSE predicts interference
whereas the exact analysis does not.

The fundamental model we consider is a 1D gas of $N$ hard core bosons
at zero temperature described by the Schr\"{o}dinger Hamiltonian
\begin{equation}\label{eq1}
\hat{H}=\sum_{j=1}^{N}-\frac{\hbar^2}{2m}\frac{\partial^2}
{\partial x_{j}^{2}}
+V(x_{1},\cdots,x_{N};t)  ,
\end{equation}
where $x_j$ is the one-dimensional position of the $j{\it th}$ particle
and the many-body potential $V$ is symmetric (invariant) under
permutations of the particles.  The two-particle interaction
potential is assumed to contain a hard core of 1D diameter $a$. This is
conveniently treated as a constraint on allowed wave functions
$\psi(x_{1},\cdots,x_{N};t)$:
\begin{equation}\label{eq2}
\psi=0\quad\text{if}\quad |x_{j}-x_{k}|<a\quad,\quad 1\le j<k\le N  ,
\end{equation}
rather than as an infinite contribution to $V$, which then consists of all
other (finite) interactions and external potentials.  To construct
time-dependent many-boson solutions of Eq. (\ref{eq1}) we employ the
Fermi-Bose mapping \cite{Gir60,GirWri00a,RojCohBer99} and start
from fermionic solutions $\psi_{F}(x_{1},\cdots,x_{N};t)$ of
the time-dependent many-body Schr\"{o}dinger equation (TDMBSE)
$\hat{H}\psi=i\hbar\partial\psi/\partial t$ which are antisymmetric under
all particle pair exchanges $x_{j}\leftrightarrow x_{k}$, hence all
permutations. Next introduce a ``unit antisymmetric function"
\begin{equation}\label{eq3}
A(x_{1},\cdots,x_{N})=\prod_{1\le j<k\le N}\text{sgn}(x_{k}-x_{j})  ,
\end{equation}
where $\text{sgn}(x)$ is the algebraic sign of the coordinate difference
$x=x_{k}-x_{j}$, i.e., it is +1(-1) if $x>0$($x<0$). For given 
antisymmetric $\psi_F$, define a bosonic wave function $\psi_B$ by
\begin{equation}\label{eq4}
\psi_{B}(x_{1},\cdots,x_{N};t)=A(x_{1},\cdots,x_{N})\psi_{F}(x_{1},\cdots,
x_{N};t)
\end{equation}
which defines the Fermi-Bose mapping. Then $\psi_B$ satisfies
the hard core constraint (2) if $\psi_F$ does, is totally
symmetric (bosonic) under permutations, and obeys the same
boundary conditions \cite{Gir60,Gir65}. In the Olshanii limit \cite{Ols98}
(low density, very thin atom waveguide, large scattering length) the dynamics
reduces to that of the impenetrable point Bose gas, the $a\rightarrow 0$
limit of Eq. (\ref{eq2}). Then under the assumption that
the many-body potential $V$ of Eq. (\ref{eq1}) is a sum of one-body
external potentials $V(x_{j},t)$, the solution of the fermion
TDMBSE can be written as a determinantal wavefunction
\cite{Gir60,RojCohBer99}
\begin{equation}\label{eq5}
\psi_{F}(x_{1},\cdots,x_{N};t)=C\det_{i,j=1}^{N}\phi_{i}(x_{j},t)  ,
\end{equation}
where
\begin{equation}\label{eq6}
i\hbar\frac{\partial\phi_i(x,t)}{\partial t} =
\left [-\frac{\hbar^2}{2m}\frac{\partial^2}
{\partial x^{2}}
+V(x,t)\right ]\phi_i(x,t)  .
\end{equation}
It then follows that $\psi_F$ satisfies the TDMBSE,
and it satisfies the impenetrability constraint (vanishing when any 
$x_{j}=x_{\ell}$) trivially due to antisymmetry. Then by the mapping theorem
$\psi_B$ of Eq.(\ref{eq4}) satisfies the same TDMBSE.
For each $i=1,\ldots,N$ the initial normalized wave functions
$\phi_i(x,0)$ represent different orbitals of the potential at
$t=0$.  For example, for a harmonic trapping potential $\phi_i(x,0)$
are given by the well-known Hermite-Gaussian orbitals, with one
particle occupying each of the $N$ lowest levels for the $N$-particle
ground state \cite{KolNewStr00}.

\begin{figure}
\epsfxsize 3in
\epsfbox{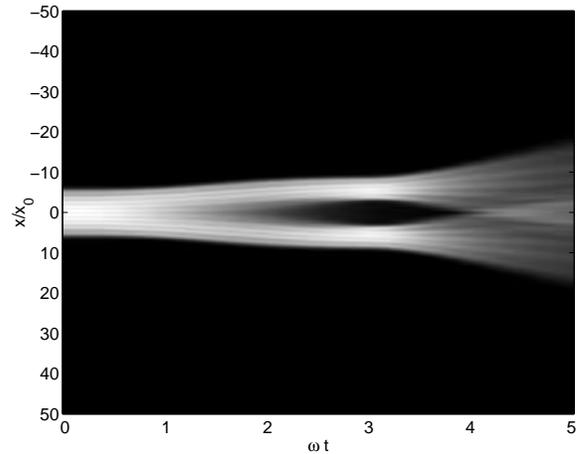}
\caption{Exact many-body theory simulation of the cool, cut, interfere
scenario. The figure shows a gray-scale plot of the particle density
$\rho_N(x,t)$ as a function of $\omega t$ (horizontal axis) and
position $x/x_0$ (vertical axis), with white being the highest density, 
for $N=10$, $V_B=20\hbar\omega$, $w=3x_0$, and $\omega t_{pot}=3$.}
\label{Fig:one}
\end{figure}

We have performed numerical simulations based on Eqs. (\ref{eq6}), using
the split-step beam propagation method \cite{FleMorFei76}, for
an initial harmonic trap containing $N$ particles in their ground state
which is subjected to a centered Gaussian repulsive potential, {\it e.g.}, 
a blue-detuned laser field which serves to split the initial state. 
After some time $t_{pot}$
both the harmonic trap and repulsive potential are turned off and the
two split components allowed to recombine: this is an interference
experiment of the cool, cut, interfere variety previously discussed
\cite{JavWil97}.  The potential in Eq. (\ref{eq6})
is taken of the specific form
\begin{equation}\label{eq7}
V(x,t) = \frac{1}{2}m\omega^2x^2 + V_B\sin(\frac{\pi t}{2t_{pot}})
e^{-x^2/w^2}  ,
\end{equation}
and $V=0$ for $t>t_{pot}$.  Figure \ref{Fig:one} shows an illustrative example
for $N=10$ with a repulsive potential of height $V_B=20\hbar\omega$, and
width $w=3x_0$, with $x_0=\sqrt{\hbar/2m\omega}$ the ground state
harmonic oscillator width, and $\omega t_{pot}=3$.
This figure shows a gray scale plot of the $N$-particle density
$\rho_N(x,t)=\sum_{i=1}^N|\phi_i(x,t)|^2$ \cite{Gir60,GirWri00a,RojCohBer99}
as functions of $\omega t$ (horizonatal axis) and position $x/x_0$
(vertical axis), with white being the highest density.
\noindent
The potential height
was chosen such that $V_B>\mu=\hbar\omega(N-1/2)$, where $\mu$ is the
chemical potential of the $N$-particle oscillator ground state
\cite{KolNewStr00}, noting that the top of the $N$-particle Fermi sea is at 
$n=N-1$. As expected, as the repulsive
potential turns on it splits the initial ground state into two separated
components.  Upon release at $t=t_{pot}$ the two components expand and
subsequently recombine.  What is noteworthy is that although there is
some modulation upon recombination there are no strong modulations
indicative of the interference provisionally expected for bosons: this
was a generic finding from our simulations irrespective of the time
scale on which the repulsive potential was turned on
\cite{JavWil97,LegSol98}.  In contrast, the density $|\phi_i(x,t)|^2$
for each individual orbital $i=1,\ldots,N$ can show large
inteference
fringes, but with a different period in each case.  Thus, the minimal
inteference seen in Fig. 1 is a result of washing out of the individual
interferences by averaging over N orbitals.  Thus, the remnant of any
interference fringes decreases with increasing $N$ and vanishes in the
thermodynamic limit.

Physically, it makes sense that the interference fringes are all but
absent since the Fermi-Bose mapping shows that in this 1D limit the
system of bosons acts effectively like a system of free fermions insofar as
effects expressible only in terms of $|\psi_{B}|^2$ are concerned, so 
interference is not expected \cite{CahGla99}.  The lack of
interference is therefore a signature of the Fermi-Bose duality that
occurs in 1D systems of impenetrable particles \cite{Gir60}.

\begin{figure}
\epsfxsize 3in
\epsfbox{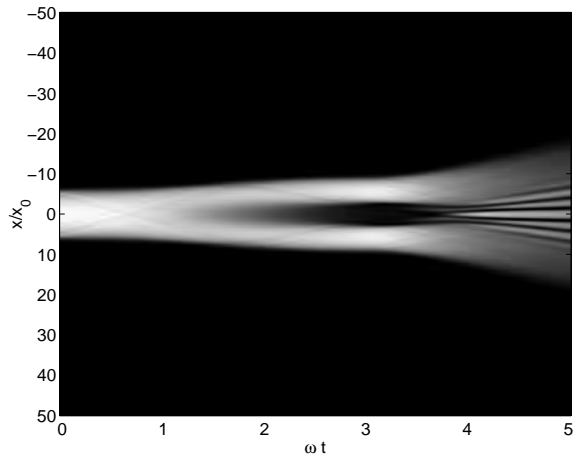}
\caption{Mean-field theory simulation of the cool, cut, interfere scenario.
The figure shows a gray-scale plot of the particle density
$\rho(x,t)=|\Phi(x,t)|^2$ as a function of $\omega t$ (horizonatal axis) and
position $x/x_0$ (vertical axis). Parameters are the same as Fig. 1.}
\label{Fig:two}
\end{figure}

We next turn to the mean-field description proposed by Kolomeisky
{\it et al.} \cite{KolNewStr00} for low-dimensional systems.  In
particular, they introduce an order parameter $\Phi(x,t)$, normalized
to the number of particles $N$, for such
systems, though they do not discuss what this represents physically.
Using energy functional arguments they deduce the following
NLSE with quartic nonlinearity for a 1D system of impenetrable
bosons:
\begin{equation}\label{eq8}
i\hbar\frac{\partial\Phi}{\partial t} =
\left [-\frac{\hbar^2}{2m}\frac{\partial^2}
{\partial x^{2}}
+V(x,t) + \frac{(\pi\hbar)^2}{2m}|\Phi|^4 \right ]\Phi  .
\end{equation}
with $V(x,t)$ the same as in Eq. (\ref{eq6}).  Our goal now is to
compare the predictions of this NLSE for the same cool, cut, interfere
simulation in Figure \ref{Fig:one}, with the initial condition
$\Phi(x,0)=\sqrt{\rho_N(x,0)}$ corresponding to the exact many-body
solution, all other parameters being equal.  Figure \ref{Fig:two} shows the
corresponding gray-scale plot of the density $\rho(x,t)=|\Phi(x,t)|^2$, and
two features are apparent: First, during the splitting phase when the
repulsive potential is on there is very good overall agreement between
the exact many-body theory and the NLSE prediction.  Second, when the
split components are released and recombine they produce pronounced
interference fringes, in contrast to the exact theory.  Indeed, this
interference in the MFT is to be expected on the basis of previous
theoretical work \cite{NarWalSch96}, even though a quartic (rather than 
quadratic) nonlinearity
is employed here.  Thus the MFT cannot accurately capture the time-dependent
dynamics in all situations.

In conclusion, the numerical simulations presented show that MFT seriously 
overestimates the magnitude of interference phenomena in a 1D gas of 
impenetrable bosons. However, Kolomeisky {\it et al.} have demonstrated that 
their NLSE produces excellent results for the ground state density
profile of 1D atomic clouds, and our simulations show that during
the splitting process there is excellent agreement between the
exact theory and the MFT.  During the splitting the density profile
is found to closely follow the ground state of the local applied
potential, harmonic trap plus repulsive potential, meaning that
the phase of the order paramter is not a key element during this time 
interval. In this case the order parameter for the NLSE is physically
$\Phi(x,t)=\sqrt{\rho(x,t)}$, to within an arbitary overall phase
factor.  In contrast, during the recombination phase the relative
phase of the order parameter between the two split components
is the key element allowing for the appearence of the interference fringes
in the MFT.  Since the exact many-body theory does not show such
fringes this means that the MFT approach endows the order parameter
with phase information that is beyond the real degree of coherence
present in the system.  Thus, as long as this coherence does not manifest
itself in the form of interference, the MFT is of some utility, but
not if the predictions are sensitive the the order parameter phase.
This view is consistent with the lack of off-diagonal-long-range-order
in the 1D gas of impenetrable bosons \cite{Len66,VTr79},
which is normally required for the introduction of an order parameter
\cite{PenOns56,Yan62}.  In future work we hope to investigate the
transition of impenetrable bosons from the 1D regime, with no interference
effects, to the 3D regime, which is well described by MFT and shows
interference \cite{NarWalSch96}. In particular, at 1D atom waveguide
junctions such as atom beam splitters or recombiners, the system
will be sensitive to the fact that the 1D waveguides are actually
embedded in a higher-dimensional space, in which case atom interferometers
employing 1D atom waveguides may still be capable of producing interference
fringes.  Another very interesting question is
whether or not the usual GP NLSE or its 2D modification proposed by
Kolomeisky {\it et al.} \cite{KolNewStr00}
gives reliable predictions of interference between BECs in the 2D regime. 
This has potentially important experimental consequences in view of progress
in construction of 2D atom waveguides and their potential applicability to
atom interferometry \cite{Ber97}.
\vspace{0.2cm}

\noindent
This work was supported by the Office of Naval Research Contract
No. N00014-99-1-0806.


\begin{references}
%
\bibitem{DalGioPit99} For a recent review see F. Dalfovo {\it et al.},
\Journal{\RMP}{71}{463}{1999}.
%
\bibitem{BayPet96} G. Baym and C. J. Pethick, \Journal{\PRL}{76}{6}{1996}.
%
\bibitem{KagShlWal96} Y. Kagan, G. V. Shlapnikov, and J. T. M. Walraven,
\Journal{\PRL}{76}{2670}{1996}.
%
\bibitem{EdwDodCla96} M. Edwards {\it et al.}, \Journal{\PRL}{77}{1671}{1996}.
%
\bibitem{Str96} S. Stringari, \Journal{\PRL}{77}{2360}{1996}.
%
\bibitem{GolPlaMey95} E. V. Goldstein, K. Pl\"attner, and P. Meystre,
{\sl Quantum. Semiclass. Opt.} {\bf 7}, 743 (1995).
%
\bibitem{DenHagWen99} L. Deng {\it et al.}, {\sl Science} {\bf 398},
218 (1999).
%
\bibitem{MorBalBur97} S. A. Morgan, R. J. Ballagh, and K. Burnett,
\Journal{\PRA}{55}{4338}{1997}.
%
\bibitem{ReiCla97} W.P. Reinhardt and C.W. Clark, \Journal{\JPB}{30}{L785}
{1997}.
%
\bibitem{NarWalSch96} M. Naraschewski {\it et al.},
\Journal{\PRA}{54}{2185}{1996}.
%
\bibitem{JavWil97} J. Javanainen and M. Wilkens,
\Journal{\PRL}{78}{4675}{1997}; {\it ibid.} {\bf 81}, 1345 (1998).
%
\bibitem{LegSol98} A. J. Legget and F. Sols,
\Journal{\PRL}{81}{1344}{1998}.
%
\bibitem{WriWonCol97} E. M. Wright {\it et al.},
\Journal{\PRA}{56}{591}{1997}.
%
\bibitem{PenOns56} O. Penrose and L. Onsager, {\sl Phys. Rev.} {\bf 104}, 576
(1956).
%
\bibitem{Yan62} C. N. Yang, \Journal{\RMP}{34}{694}{1962}.
%
\bibitem{Gir60} M. Girardeau, \Journal{\JMP}{1}{516}{1960}.
%
\bibitem{Gir65} M.D. Girardeau, \Journal{\PR}{139}{B500}{1965}.
See particularly Secs. 2, 3, and 6.
%
\bibitem{Len66} A. Lenard, \Journal{\JMP}{7}{1268}{1966}.
%
\bibitem{VTr79} H.G. Vaidya and C.A. Tracey, \Journal{\PRL}{42}{3}{1979}.
%
\bibitem{Ols98} M. Olshanii, \Journal{\PRL}{81}{938}{1998}.
%
\bibitem{HinBosHug98} E. A. Hinds {\it et al.},
\Journal{\PRL}{80}{645}{1998}; J. Schmiedmayer, \Journal{\EPJD}{4}{57}{1998}; 
M. Key {\it et al.}, \Journal{\PRL}{84}{1371}{2000}; D. M\"{u}ller 
{\it et al.}, physics/9908031.
%
\bibitem{ThyWesPre99} J. H. Thywissen {\it et al.}, \Journal{\EPJD}{4}{57}
{1998}; \Journal{\PRL}{83}{3762}{1999}; \Journal{\EPJD}{7}{261}{1999}.
%
\bibitem{GirWri00a} M. D. Girardeau and E. M. Wright, ``Many-body
solitons in a one-dimensional condensate of hard core bosons,"
cond-mat/0002062.
%
\bibitem{RojCohBer99} A. G. Rojo, G. L. Cohen, and P. R. Berman,
\Journal{\PRA}{60}{1482}{1999}.
%
\bibitem{KolNewStr00} E. B. Kolomeisky, T. J. Newman, J. P. Straley,
and Xiaoya Qi, ``Low-dimensional Bose liquids: Beyond the Gross-Pitaevskii
equation," cond-mat/0002282. Note that the expression for the condensate
width given following Eq. (8) therein is in error; the factor $m\omega$
should be in the denominator of the square root instead of the numerator.
%
\bibitem{FleMorFei76} J. A. Fleck, J. R. Morris, and M. D. Feit,
{\sl Appl. Opt.} {\bf 10}, 129 (1976).
%
\bibitem{CahGla99} K. E. Cahill and R. J. Glauber,
\Journal{\PRA}{59}{1538}{1999}.
%
\bibitem{Ber97} {\it Atom Interferometry}, ed. P. R. Berman (Academic Press,
Boston, 1997).
\end{references}
\end{document}